\documentclass[twocolumn,aps,prl,nobalancelastpage]{revtex4-1}
\usepackage{multirow}
\usepackage{graphicx}
\usepackage{tabularx}
\usepackage{booktabs}
\usepackage{amssymb}
\usepackage{amsmath}
\usepackage[colorlinks,urlcolor=blue,citecolor=blue,linkcolor=blue]{hyperref}

\usepackage{epsfig,graphicx}%[draft]
\usepackage{flafter}
\usepackage{color}
\usepackage{braket}

\begin{document}

%\preprint{}

\title{Second sound and the superfluid fraction in a resonantly interacting Fermi gas} %Title of paper



%\author{Meng Khoon Tey}
%\thanks{}
\affiliation{}

\author{Leonid A. Sidorenkov}
\author{Meng Khoon Tey}
\author{Rudolf Grimm}
\affiliation{Institut f\"ur Quantenoptik und Quanteninformation (IQOQI),
 \"Osterreichische Akademie der Wissenschaften}
 \affiliation{Institut f\"ur Experimentalphysik, Universit\"at Innsbruck, 6020 Innsbruck, Austria}

\author{Yan-Hua Hou$^{1}$}
\author{Lev Pitaevskii$^{1,2}$}
\author{Sandro Stringari$^{1}$}
\affiliation{$^1$Dipartimento di Fisica, Universit\'a di Trento and INO-CNR BEC Center, I-38123 Povo, Italy}
 \affiliation{$^2$Kapitza Institute for Physical Problems RAS, Kosygina 2, 119334 Moscow, Russia}


\date{\today}

\begin{abstract}
Superfluidity is a macroscopic quantum phenomenon, which shows up below a critical temperature \cite{Kapitza1938vol, Allen1938fpi} and leads to a peculiar behavior of matter, with frictionless flow, the formation of quantized vortices, and the quenching of the moment of inertia being intriguing examples. A remarkable explanation for many phenomena exhibited by a superfluid at finite temperature can be given in terms of a two-fluid mixture \cite{Tisza1938tpi,Landau1941tto} comprised of a normal component that behaves like a usual fluid and a superfluid component with zero viscosity and zero entropy. Important examples of superfluid systems are liquid helium and neutron stars.
More recently, ultracold atomic gases have emerged as new superfluid systems with unprecedented possibilities to control interactions and external confinement. Here we report the first observation of `second sound' in an ultracold Fermi gas with resonant interactions. Second sound is a striking manifestation of the two-component nature of a superfluid and corresponds to an entropy wave, where the superfluid and the non-superfluid components oscillate in opposite phase, different from ordinary sound (`first sound'), where they oscillate in phase.
The speed of second sound depends explicitly on the value of the superfluid fraction \cite{Khalatnikov1965book}, a quantity sensitive to the spectrum of elementary excitations \cite{Landau1947ott}. Our measurements allow us to extract the temperature dependence of the superfluid fraction, which in strongly interacting quantum gases has been an inaccessible quantity so far.
\end{abstract}

%\pacs{}% insert suggested PACS numbers in braces on next line

\maketitle %\maketitle must follow title, authors, abstract and \pacs

% References ishould be done using the \cite, \ref, and \label commands

Until now, second sound has only been observed in superfluid helium \cite{Atkins1959book}, the paradigm of quantum fluids characterized by strong interactions, for the description of which Landau developed his theory of two-fluid hydrodynamics \cite{Landau1941tto}. In liquid helium, second sound can be generated \cite{Peshkov1944ssi} by local time-dependent heating and detected by observing the propagation of the resulting temperature wave. Second sound is an essentially isobaric oscillation, in contrast to the adiabatic nature of first sound.

In experiments with ultracold atomic quantum gases, the observation of second sound has been a long-standing, but so far elusive goal. In dilute Bose-Einstein condensed samples, the relative motion of the condensate with respect to the thermal component represents a related effect, the observation of which has been reported in refs.~\cite{Stamperkurn1998cah, Meppelink2009dos}. However, a direct observation of second-sound waves has not been achieved.
In resonantly interacting Fermi gases \cite{Giorgini2008tou, Bloch2008mbp}, superfluidity and the universal thermodynamics \cite{Ho2004uto,Kinast2005hco,Horikoshi2010mou,Nascimbene2010ett, Ku2012rts} have been subjects of intense research. In these systems also the normal component behaves in a deeply hydrodynamic way, which means that Landau's two-fluid theory can be readily applied. This suggests a behavior similar to superfluid helium, including the occurrence of second sound.

Our system is an ultracold, superfluid sample of fermionic $^6$Li atoms, prepared in a highly elongated harmonic trapping potential (Methods) by well-established procedures of laser and evaporative cooling \cite{Tey2013cmi}. The sample consists of $N = 3.0 \times 10^5$ atoms in a balanced mixture of the two lowest spin states, and is about $500\,\mu$m long and $20\,\mu$m wide. It is characterized by the Fermi temperature $T_F^{\rm trap} \approx 0.9\,\mu$K (Methods). A magnetic bias field of 834\,G is applied, which tunes the interaction between the two spin components right on top of an $s$-wave scattering length resonance \cite{Chin2010fri} (`unitarity limit' of interactions). The cloud's temperature $T$ is determined by analyzing axial density profiles \cite{Nascimbene2010ett, Tey2013cmi}, using the knowledge of the equation of state (EOS) from ref.~\cite{Ku2012rts}.

\begin{figure}[t]
\begin{center}
\includegraphics[width=0.8\columnwidth]{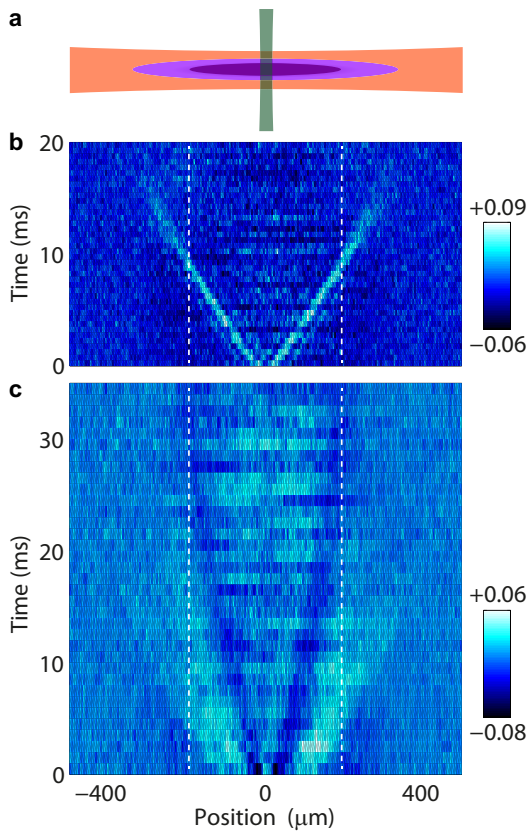}
\caption{\textbf{Observing the propagation of first and second sound.} \textbf{a}, The basic geometry of exciting the optically trapped cloud with a weak, power-modulated repulsive laser beam (green), which perpendicularly intersects the trapping beam (red). The trapped cloud has a superfluid core, surrounded by a normal region. \textbf{b} and \textbf{c}, Normalized differential axial density profiles $\delta n_1(z, t)/n_{1, \, {\rm max}}$ taken for variable delay times after the excitation show the propagation of first sound (local density increase, bright) and second sound (local decrease, dark). The temperature of the atomic cloud is $T = 0.135(10) T_F^{\rm trap}$. The vertical dashed lines indicate the axial region where superfluid is expected to exist according to a recent determination of the critical temperature \cite{Ku2012rts}.}
\label{fig:soundprop}
\end{center}
\end{figure}

Our method for observing sound propagation builds on the classical scheme to detect the propagation of first sound in Bose-Einstein condensates \cite{Andrews1997pos}, which was also applied to resonantly interacting Fermi gases \cite{Joseph2007mos}.
The general idea is to prepare the quantum gas in a trap that is highly elongated, to create a local perturbation, and to detect its one-dimensional propagation as a pulse along the long trap axis. In the case of first sound, both the creation and the detection of such an excitation are straightforward, while being less obvious for second sound.

For the local excitation of the cloud, we use a repulsive dipole potential created by a tightly focussed green laser beam (Methods) that perpendicularly intercepts the trapped sample in its center, as shown in Fig.~\ref{fig:soundprop}a. To excite first sound, we suddenly turn on the repulsive beam. The local reduction of the trapping potential acts on the superfluid and normal component in the same way and creates a small hump in the axial density distribution.
To excite second sound, we keep the green beam's power constant during the whole experimental sequence with the exception of a short power-modulation burst, which contains 8 sinusoidal oscillations in 4.5 ms (Methods). The fast modulation locally drives the system out of equilibrium and the following relaxation increases entropy and temperature. The duration of the burst is chosen such that the system can establish a local thermal equilibrium on a length scale that covers the transverse cloud size, but is much shorter than the axial extension of the cloud. In all cases, we take care that the excitation remains a small perturbation of the whole system, which globally stays in a thermal equilibrium state.

For detection we record the axial density profile $n_1(z,t)$ for various time delays $t$ after the excitation pulse, where $n_1(z,t)$ is the number density integrated over the transverse degrees of freedom. To enhance the visibility of the density perturbation, we subtract a background profile $\bar{n}_1(z)$ obtained by averaging the profiles over all measured delay times. Our signal $\delta n_1(z, t) = n_1(z, t) - \bar{n}_1(z)$ is finally normalized to the maximum observed density
$n_{1, \, {\rm max}}$.

The key point for the detection of second sound is the coupling \cite{Arahata2009pos,Hu2010ssa} between temperature and density variations, which occurs in a systems exhibiting thermal expansion. The relevant isobaric thermal expansion coefficient can be obtained from the EOS and, for our experimental conditions, is found to be sufficiently large to facilitate the observation of a local temperature increase as a dip in the density profiles (Methods).

First sound clearly shows up in Fig.~\ref{fig:soundprop}b. The initially induced hump splits into two density peaks (bright), which symmetrically propagate outward at an almost constant speed, penetrate into the region where there is no superfluid (see dashed lines for the superfluid-normal boundary), and finally fade out in the outer region of the cloud. For longer times, we observe (not shown) a weak collective breathing oscillation to be excited.

\begin{figure}[t]
\includegraphics[width=1\columnwidth]{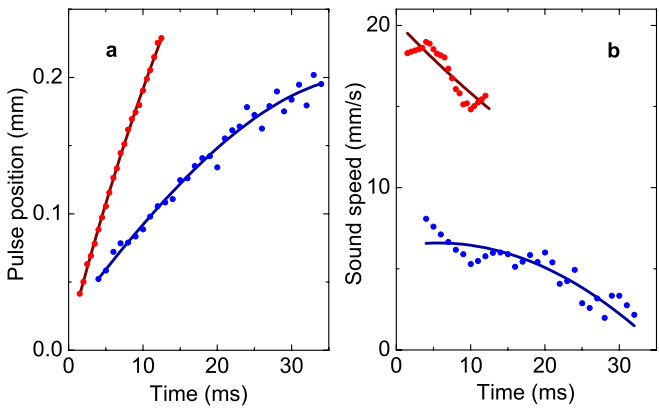}
\caption{ \textbf{Extracting the sound speeds.} \textbf{a}, The positions of the propagating pulses are shown as a function of time. The data points (red and blue symbols for first and second sound, respectively) result from individual fits to the pulses observed at fixed delay times, and the solid lines represent third-order polynomial fits to the time-dependent behavior. \textbf{b}, The sound speeds are obtained as derivatives of the fit curves (solid lines) and alternatively by analyzing subsets of nine adjacent profiles (data points).}
\label{fig:dataanalysis}
\end{figure}

The excitation with our local heating scheme leads to a strikingly different picture, as shown in Fig.~\ref{fig:soundprop}c. The two density dips (dark) propagate much slower than the first-sound signals. They further slow down when approaching the superfluid boundary (dashed lines) and finally disappear without penetrating into non-superfluid region. This behavior is our `smoking gun' of second sound.

To extract the two sound speeds from the differential profiles $\delta n_1(z, t)$, we determine the positions of the density dips or peaks with Gaussian fit functions. The corresponding time dependence is shown in Fig.~\ref{fig:dataanalysis}a, as extracted from the profiles in Fig.~\ref{fig:soundprop}. We now use a third-order polynomial
%$z_\mathrm{pulse}(t) = c_0+c_1t+c_2t^2+c_3t^3$
to globally fit the time-dependent positions (solid lines in Fig.~\ref{fig:dataanalysis}a). The sound speeds are then obtained as time derivatives of the fit curves and displayed as solid lines in Fig.~\ref{fig:dataanalysis}b. As this procedure by design produces a smooth curve and does not provide sufficient insight into the uncertainties, we also apply a different procedure to analyze the same data. We consider smaller subsets of adjacent points and extract the local speeds by second-order polynomial fits. Corresponding results are shown in Fig.~\ref{fig:dataanalysis} by the filled symbols.

The fact that the axial harmonic confinement introduces a $z$-dependence of the linear density $n_1$ allows us to determine the temperature dependence of the sound speeds without changing the global temperature $T$ of the trapped sample. The key is to define a $z$-dependent Fermi temperature $T_F^{1D} \propto n_1^{2/5}$ (Methods) as the natural local temperature scale.
The corresponding reduced temperature $T/T_F^{1D}$ has its minimum at the trap center ($z=0$) and increases with $z$. The superfluid phase-transition is crossed when the critical temperature $T_c = 0.214(16) T_F^{\rm 1D}$ is reached.
\begin{figure}[t]
\begin{center}
\includegraphics[width=0.85\columnwidth]{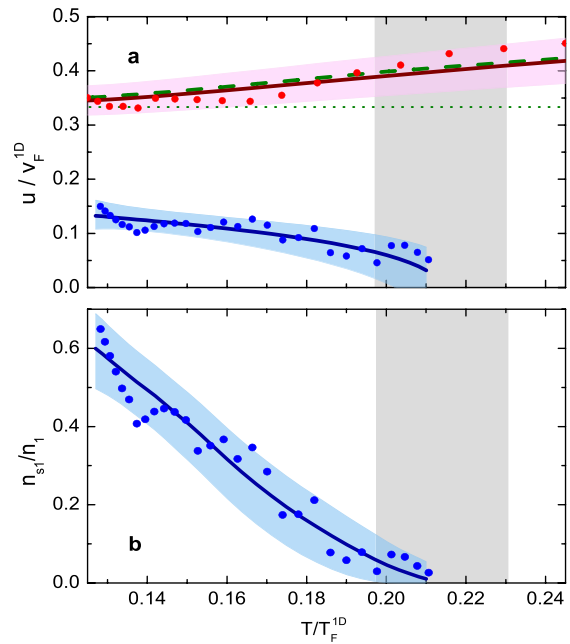}
\caption{ \textbf{Normalized sound speeds and the 1D superfluid fraction.} \textbf{a}, Speeds of first and second sound, normalized to the local Fermi speed and plotted as a function of the reduced temperature. The data points and the solid lines refer to the data set of Fig.~\ref{fig:soundprop}, following different methods to analyze the raw data (see text). The shaded regions indicate the maximum range of variations from analyzing different data sets. The dashed curve is a prediction based on Eq.~\ref{eq:u1} and the EOS from ref.~\cite{Ku2012rts}. The dotted horizontal line is the corresponding zero-temperature limit \cite{Capuzzi2006spi} for the speed of first sound.
\textbf{b}, Temperature dependence of the 1D superfluid fraction $n_{s1}/n_1$, with symbols, solid line, and shaded uncertainty range corresponding to panel \textbf{a}. In both panels, the grey shaded area indicates the uncertainty range of the superfluid phase transition according to ref.~\cite{Ku2012rts}.}
\label{fig:soundspeed}
\end{center}
\end{figure}

In Fig.~\ref{fig:soundspeed}, we show the temperature dependence of the two speeds of sound, normalized to the local Fermi speed $v^{\rm 1D}_F = \sqrt{2 k_B T^{\rm 1D}_F /m}$\,, where $m$ is the atomic mass and $k_B$ is Boltzmann's constant. The symbols correspond to the data displayed in Fig.~\ref{fig:dataanalysis}. The solid lines are derived in the same way from the corresponding fit curves. To get additional information on the confidence level of our results, we have analyzed a number of data sets taken under similar conditions as the ones in Fig.~\ref{fig:soundprop}. The regions shaded in pink and light-blue display the maximum range of variations considering all our different data sets.

%The excitation of sound and its subsequent experimental detection are favored by this 1D geometry and theory  takes a particularly simple form, allowing for a direct interpretation of the experimental results.

Our interpretation of the experimental results relies on an effective one-dimensional approach that allows to solve Landau's two-fluid hydrodynamic equations for a highly elongated system \cite{Bertaina2010fas, Hou2013fas}. The basic assumptions are a thermal equilibrium in the radial direction and sufficient shear viscosity to establish a flow field that is independent of the radial position. Within this theoretical framework and under the local density approximation, effective 1D thermodynamic quantities can be defined by integration over the transverse degrees of freedom, such that a thermodynamic quantity $q$ yields a 1D counterpart $q_1 \equiv 2\pi \int_0^\infty q \, r \, dr$.
We can express the normalized speeds of first and second sound as
\begin{equation}
\frac{u_1}{v^{\rm 1D}_F}= \sqrt{\frac{7}{10}\frac{P_1}{n_1 k_B T^{\rm 1D}_F}} \, ,
\label{eq:u1}
\end{equation}
and
%\begin{equation}
%\frac{u_2}{v^{\rm 1D}_F} = \sqrt{\frac{T}{2 k_B T^{1D}_F}\frac{s_1^2}{n_1 C_{p1}}\frac{n_{s1}}{n_{n1}}}, %\tag{2}
%\label{eq:u2}
%\end{equation}
\begin{equation}
\frac{u_2}{v^{\rm 1D}_F} = \sqrt{\frac{T}{2 k_B T^{1D}_F}\frac{\bar{s}_1^2}{\bar{c}_{p1}}\frac{n_{s1}}{n_{n1}}},
\label{eq:u2}
\end{equation}
where $P_1$ denotes the 1D pressure (unit of a force), $\bar{s}_1 = s_1/n_1$ is the entropy per particle, and $\bar{c}_{p1} = T (\partial \bar{s}_1 / \partial T)_{p1}$ is the isobaric heat capacity per particle. These thermodynamic quantities can be calculated from the EOS as functions of the reduced temperature $T/T_F^{1D}$, as we discuss in detail in ref.~\cite{Hou2013fas}. The quantities $n_{s1}$ and $n_{n1}=n_1 - n_{s1}$, which represent the linear number densities of the superfluid and the normal component, cannot be determined from the known EOS.

The speed of first sound provides us with an important benchmark for our experimental method and the interpretation of the measurements in the 1D theoretical framework. The experimental results (red symbols and upper solid line in Fig.~\ref{fig:soundspeed}a) are in excellent agreement with the calculation (dashed line) based on Eq.~\ref{eq:u1} and the EOS from ref.~\cite{Ku2012rts}, which is a further confirmation for the validity of our theoretical approach in addition to the recent measurements of the $T$-dependent frequencies of higher-nodal collective modes \cite{Tey2013cmi}.

\begin{figure}[t]
\begin{center}
\includegraphics[width=0.85\columnwidth]{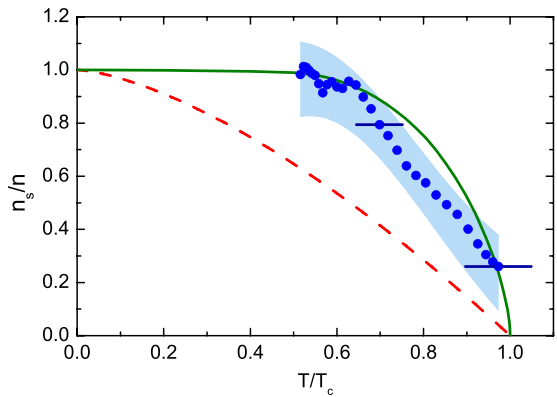}
\caption{
\textbf{Superfluid fraction for the homogeneous case.} The data points and the corresponding uncertainty range (shaded region) show the superfluid fraction for a uniform resonantly interacting Fermi gas versus $T/T_c$ as reconstructed from its 1D counterpart in Fig.~\ref{fig:soundspeed}b (Methods). The two horizontal error bars indicate the systematic uncertainties resulting from the limited knowledge of the critical temperature $T_c$. For comparison, we show the fraction for helium II (solid line) as measured in ref.~\cite{Dash1957hoo} and the textbook expression $1-(T/T_c)^{3/2}$ for the Bose-Einstein condensed fraction of the ideal Bose gas (dashed line).}
\label{fig:uniform}
\end{center}
\end{figure}

The measured speed of second sound (blue line and data points in Fig.~\ref{fig:soundspeed}a) is observed to decrease with temperature, in contrast to first sound. The general behavior fits very well to the qualitative predictions in ref.~\cite{Bertaina2010fas}. We can now extract the 1D superfluid fraction $n_{s1}/n_1$, which is the unknown quantity in Eq.~\ref{eq:u2}. The result is presented in Fig.~\ref{fig:soundspeed}b, where $n_{s1}/n_1$ shows a smooth increase with decreasing $T$ below the critical temperature.

We finally reconstruct (Methods) the temperature dependence of the superfluid fraction $n_s/n$ for the homogeneous 3D case, which has been an inaccessible quantity so far. The results, shown in Fig.~\ref{fig:uniform}, turn out to be rather close to the case of liquid helium II (solid line) \cite{Dash1957hoo}. In particular, the superfluid part is practically 100\% below $0.6\,T_c$. This behavior is quite different from the one exhibited by a weakly interacting Bose gas, whose superfluid fraction lies significantly below the data of Fig.~\ref{fig:uniform} and is well approximated by the condensate fraction of the ideal Bose gas (dashed line). %$1-(T/T_c)^{3/2}$.
In strongly interacting quantum fluids, the superfluid and the condensate fractions simultaneously appear at the phase transition, but they exhibit quite different temperature dependencies below $T_c$. Our experimental results provide a new benchmark for advancing theoretical approaches to calculate the superfluid fraction, which is a challenging problem in quantum many-body physics.

From the fundamental point of view, the experimentally determined superfluid fraction represents a so far missing thermodynamic function, which contains information on the spectrum of elementary excitations and completes the description of the superfluid in terms of universal thermodynamics. With respect to future applications, the creation of second sound represents a paradigm for controlling the relative motion of a superfluid with respect to the normal component. This may find applications in other situations and geometries, where resonantly interacting Fermi gases are used as model systems for exploring dynamical and transport phenomena \cite{Nascimbene2009coo,Sommer2011ust,Stadler2012otd}.

We thank E.\ R.\ S\'anchez Guajardo for his contributions in the early stage of this work. The Innsbruck team acknowledges support from the Austrian Science Fund (FWF) within SFB FoQuS (project No.\ F4004-N16). The Trento team acknowledges support from the European Research Council through the project QGBE.

%\textbf{Competing Interests} The authors declare that they have no competing financial interests.

%\textbf{Author Contributions} L.A.S. and M.K.T. equally contributed to the experimental work and the data analysis under the supervision of R.G. The new experimental methods were conceived by these three authors jointly. The theoretical work was performed by Y.H.H., L.P., S.S.

%\textbf{Correspondence} Correspondence and requests for materials should be addressed to M.K.T.~(email: mengkhoon.tey@ultracold.at).

\section{Methods}

{\bf Trapping potential.} In our hybrid trap \cite{Jochim2003bec}, the tight radial confinement with a trapping frequency $\omega_r/(2\pi) = 539(2)$\,Hz is provided by an infrared laser beam (wavelength 1075\,nm, power 120\,mW, waist $39\mu$m). The much weaker axial confinement with a trapping frequency $\omega_z/(2\pi) = 22.46(7)$\,Hz results from the curvature of the applied magnetic field. The frequency ratio $\omega_r/\omega_z \approx 24$ corresponds to the aspect ratio of the trapped cloud.

\noindent
{\bf Sound excitation and detection.} The green laser beam (wavelength 532\,nm, power $<20$\,mW) used for excitation is focused to a waist in the range of 25-35\,$\mu$m. To excite first sound, we suddenly turn it on to introduce a repulsive potential hill with a height of about 10\% of the cloud's chemical potential in the trap center. For the case of second sound, the repulsive beam is permanently on during the preparation of the quantum gas with a barrier height of about 15\% of the cloud's chemical potential. The excitation is then induced by a burst of 8 cycles of sinusoidal power modulation. The modulation frequency is set to 1720\,Hz, but the scheme is rather robust and was found to work best in a frequency range between roughly $2$ and $3.5$ times the radial trapping frequency. The burst envelope is rectangular and the modulated barrier has a peak-to-peak amplitude of 30\% of the cloud's chemical potential. After the burst, the power is set back to its initial constant value. On the time scale of the axial motion, the time-averaged power of the green beam is constant, which avoids direct excitation of first sound.

For the detection of the propagating second-sound signal, the corresponding density dip is essential. The 1D formulation of the EOS in ref.~\cite{Hou2013fas} allows us to relate the observed depth to the relative temperature change. In the temperature range of our experiments, the corresponding thermodynamic coefficient $(\delta n_1 / n_1)/(\delta T/ T)$ takes values \cite{Hou2013fas} between $-0.4$ and $-0.6$, which means that the typical 3\% relative depth of the density dip roughly corresponds to a local temperature increase of about 6\%.

{\bf 1D Fermi temperature and critical temperature.} Three different definitions for Fermi temperatures are related to natural energy scales of our trapping geometry. The homogeneous case with a 3D number density $n$ (including both spin states) is given by $k_B T_F = (3 \pi^2)^{2/3} \frac{\hbar^2}{2 m} n^{2/3}$. Within the local density approximation the corresponding Fermi energy for $N$ atoms in a three-dimensional harmonic potential is given by $k_B T^{\rm trap}_F = \hbar (3 N \, \omega^2_r \omega_z)^{1/3}$, commonly used to describe the global situation of a 3D trap. The Fermi energy of cylindrically confined cloud in the center of a 2D harmonic trap, which we refer to as `1D Fermi temperature', follows from $k_B T^{\rm 1D}_F = (\frac{15 \pi}{8})^{2/5} (\hbar \omega_r)^{4/5} (\frac{\hbar^2 n_1^2}{2 m})^{1/5}$, where the relevant density is the linear number density $n_1$. For the additional axial confinement in our trap geometry, $n_1$ and thus $T^{\rm 1D}_F$ become $z$-dependent quantities.

For specifying the critical temperature $T_c$ in units of the relevant Fermi temperature, we also distinguish between the three different situations of a homogeneous system, a 3D harmonic trap, and a cylinder with 2D harmonic confinement. For the homogeneous system, $T_c = 0.167(13)\,T_F$ was measured in ref.~\cite{Ku2012rts}. Based on the local density approximation and the experimentally determined EOS, this result can be translated into corresponding conditions for the other two situations. While for the 3D trap, $T_c = 0.223(15) T^{\rm trap}_F$ is relevant for the occurrence of the phase transition at the center of the trap, the condition $T_c = 0.214(16) T^{\rm 1D}_F$ applies to the 2D confined case. In our highly elongated trap geometry with weak axial confinement, $T^{\rm 1D}_F$ and thus $T_c$ become $z$-dependent. For a fixed global temperature $T$, the condition $T<T_c(z)$ then determines the axial range, where a superfluid exists (see illustration in Fig.~\ref{fig:soundprop}a and dashed lines in \ref{fig:soundprop}b and \ref{fig:soundprop}c).

\noindent
{\bf Reconstruction of superfluid fraction.} Within the framework of universal thermodynamics \cite{Ho2004uto}, the number density of a uniform, resonantly interacting Fermi gas can be expressed in terms of a dimensionless universal function $f_n(x)$ by $n(x,T) = \lambda_T^{-3}f_n(x)$. Here $\lambda_T =(\frac{2\pi\hbar^2}{m k_BT})^{1/2}$ is the thermal de Broglie wavelength, and the dimensionless parameter $x = \mu/k_BT$ gives the ratio between the chemical potential $\mu$ and the thermal energy $k_BT$, with a unique correspondence existing between $x$ and $T/T_F$. The function $f_n(x)$ is known from measurements of the EOS \cite{Kinast2005hco,Horikoshi2010mou,Nascimbene2010ett,Ku2012rts}. The superfluid density can be expressed in an analogous way as $n_s = \lambda_T^{-3}f_{ns}(x)$, introducing a corresponding universal function $f_{ns}(x)$, which is to be extracted from our measurements. Using the local density approximation, one can show for a system with radial harmonic confinement that \cite{Nascimbene2010ett} $n_1(x_0,T) = \frac{2\pi}{m\omega_r^2}\frac{k_B T}{\lambda_T^3}\int_{-\infty}^{x_0}f_n(x)dx$, where $x_0$ represents the value of $x$ on the trap axis. Analogously, the 1D superfluid density is given by $n_{s1}(x_0,T) = \frac{2\pi}{m\omega_r^2}\frac{k_B T}{\lambda_T^3}\int_{-\infty}^{x_0}f_{ns}(x)dx$. One thus easily sees that the 1D superfluid fraction is given by $n_{s1}/n_1 = \int_{-\infty}^{x_0}f_{ns}(x)dx/\int_{-\infty}^{x_0}f_n(x)dx$ and only depends on $x_0$. From our experimentally determined $n_{s1}/n_1$, we readily obtain the superfluid fraction of a uniform gas using the relation $\frac{n_s}{n} = \frac{f_{ns}(x_0)}{f_n(x_0)}= \frac{1}{f_n(x_0)}\frac{d}{dx_0}\left[\frac{n_{s1}}{n_1}\int_{-\infty}^{x_0}f_n(x)dx\right]$.


%\bibliography{ultracold,secondsound}
\end{document}